\documentclass{ws-mpla}
\usepackage[super]{cite}
\usepackage[utf8]{inputenc}
\usepackage{array}
\usepackage{amssymb}
\usepackage{amsmath}
\usepackage{graphicx}
\usepackage{bigstrut}
\begin{document}
\newcommand{\el}{\mbox{e$^{-}$}}
\newcommand{\gam}{\mbox{$\gamma$}}
\newcommand{\pa}[1]{\mbox{#1}}
\newcommand{\pb}[1]{\mbox{$\overline{\mathrm{#1}}$}}
\newcommand{\pac}[2]{\mbox{#1$^{#2}$}}
\newcommand{\pos}{\mbox{e$^{+}$}}
\newcommand{\half}{\mbox{$\frac{1}{2}$}}
\newcommand{\mrm}{\mathrm}
\newcommand{\ol}{\overline}
\newcommand{\ul}{\underline}
\newcommand{\ra}{\mbox{$\rightarrow$}}
\newcommand{\ri}{{\rm i}}
\newcommand\subtitle[1] {~\\ \noindent{\bf #1}\\[2mm]}

\markboth{Dezső Horváth and Zoltán Trócsányi}{Particles and antiparticles}


\title{Particles and antiparticles}

\author{Dezső Horváth}
\address{Wigner Research Centre for Physics, H-1525 Budapest, P.O.B. 49, Hungary\\
and  University Babe\textcommabelow{s}--Bolyai, Cluj-Napoca, Romania\\
horvath.dezso@wigner.hu}

\author{Zoltán Trócsányi}
\address{Institute for Theoretical Physics, ELTE E{\"o}tv{\"o}s Lor\'and University,
P\'azm\'any P{\'e}ter s{\'e}t\'any 1/A, H-1117 Budapest,
Hungary}

\maketitle

\pub{Received (Day Month Year)}{Revised (Day Month Year)}

\begin{abstract}
We review the concept of charge and chirality for particles and
antiparticles.  We point out that the commonly accepted equivalence of
particles and antiparticles — with difference only in the opposite
signs of their charges, which follows from the $CPT$ invariance — is
valid only for free non-chiral particles. We point out that with the weak
interaction turned on the equivalence of particles and antiparticles
is violated.  We also discuss that within the standard model even free
neutrinos are exceptions, they can be produced only in chiral states.
We conclude that in spite of a long history of antiparticles,
interesting theoretical and experimental challenges remain in their
complete understanding.

\end{abstract}

\ccode{PACS Nos.: 03.65.Ta, 11.30.Er, 14.60.Cd, 14.60.Pq} 

\section{Introduction}

The theory, experiments, and applications related to antiparticles are
in the centre of attention since Dirac's theoretical observation
\cite{Dirac:1928hu} of their possible existence. One could assume that
since almost immediately after Dirac's theoretical prediction
antiparticles were observed in cosmic rays, their properties are fully
described by the standard model of particle physics and confirmed by
the experimental data~\cite{ParticleDataGroup:2022pth}, there cannot
any open questions left.  In this paper we overview the
theoretical problems connected with antiparticles, with special
emphasis on concepts and misconceptions.

\section{Dirac equation}

One could doubt the efficiency of beauty in science
\cite{Hossenfelder:2018jew}, but actually, we generally adore
symmetries, and their role and contributions are very important not
only in art but in the natural sciences as
well~\cite{Horvath:2019atn}, from their origins in biology and
chemistry, through solid state physics down in scale to the micro
world. {\em Paul Dirac} did not like {\em Erwin Schrödinger's}
quantum-mechanical equation of motion, because it was not
relativistic, it treated energy and time separated from momentum and
space, thereby contradicting Einstein's unified space-time concept. He
first found a squared equation for the electron, but discarded
it\footnote{It was re-discovered later as the Klein-Gordon equation of
bosons.}, and extracted a square root of it. This is nicely described
by a frequently quoted statement of {\em Richard Feynman} ``When I was
a young man, Dirac was my hero. He made a breakthrough, a new method
of doing physics. He had the courage to simply guess at the form of an
equation, the equation we now call the Dirac equation, and to try to
interpret it afterwards.''.

Thus in 1928 Dirac published a linear equation which described the
electron with its own intrinsic angular momentum, the spin. However,
in addition to the two solutions for the spin states, there were two
more, somewhat absurd-looking solutions: electrons with positive
charge and negative energy (i.e. mass) in the two spin states. As
negative energy or mass does not exist, Dirac interpreted these
solutions as electron holes in a filled negative energy electron sea,
where electrons obey Pauli's exclusion principle, often referred to as
Dirac sea.  Considering the infinite negative charge involved, he
reconsidered this picture quite soon. Five years later {\em Carl
Anderson} observed \cite{Anderson:1933mb} positively charged
electrons (now called positrons) in cosmic rays, which had to be the
electron's antiparticles. It was gradually clarified that our
elementary material particles, the basic fermions
(Table~\ref{tab:fermions}),
and all of the composite particles,
hadrons, formed by the quarks, the basic fermions with strong
interaction, have antiparticles with the same properties, but opposite
charges. Of course, the charge distinguishing particles from
antiparticles can be other than electric as, e.g., quarks have
colours, the charges of strong interaction, whereas the antiquarks have
anticolours.

\begin{table}[tp]
\begin{center}
\tbl{Leptons and quarks, the three families of the basic fermions. 
$T_3$ is the third component of the weak isospin, the quantum number
identifying upper and lower fermions. Weak interaction groups the
fermions into left-handed doublets, and having different eigenstates,
identifies the quark flavours differently from the strong interaction,
that is denoted by the primes.}
{\begin{tabular}{|l|c|c|c|c|cc|}
\hline\hline
& Family 1 & Family 2 & Family 3 & Charge & $T_3$&\multicolumn{1}{c|}{}\bigstrut\\
\hline\hline
 &&&&&&\multicolumn{1}{c|}{}\\
Leptons&
$\left(\begin{array}{l} \nu_{\mathrm e} \\ \\ {\mathrm e} \end{array}
\right)_{\!\!\!L}$ &
$\left(\begin{array}{l} \nu_{\mu} \\ \\ \mu \end{array} \right)_{\!\!\!L}$ &
$\left(\begin{array}{l} \nu_{\tau} \\ \\ \tau \end{array} \right)_{\!\!\!L}$
& $\begin{array}{r} ~0 \\ \\  -1 \end{array} $
& $\begin{array}{r} +\frac{1}{2} \\ \\  -\frac{1}{2} \end{array} $&\multicolumn{1}{c|}{}\bigstrut\\
 &&&&&&\multicolumn{1}{c|}{}\\[-3pt]
\hline
 &&&&&&\multicolumn{1}{c|}{}\\
Quarks &
$\left(\begin{array}{l} {\mathrm u} \\ \\ {\mathrm d^{\prime}} \end{array}
\right)_{\!\!\!L} $ &
$\left(\begin{array}{l} {\mathrm c} \\ \\ {\mathrm s^{\prime}} \end{array}
\right)_{\!\!\!L} $ &
$\left(\begin{array}{l} {\mathrm t} \\ \\ {\mathrm b^{\prime}} \end{array}
\right)_{\!\!\!L}$
&
$\begin{array}{r} +\frac{2}{3} \\ \\ -\frac{1}{3} \end{array} $
& $\begin{array}{r} +\frac{1}{2} \\ \\ -\frac{1}{2} \end{array}
$&\multicolumn{1}{c|}{}\\
 &&&&&&\multicolumn{1}{c|}{}\\
\hline\hline
\end{tabular}
\label{tab:fermions}
}
\end{center}
\end{table}

The related Nobel Prizes had an interesting logic: Schrödinger and
Dirac were awarded together in 1933, after the discovery of the positron, and
Anderson himself was so three years later.

\section{Antiparticles do exist}

As they are the negative-energy solutions of the Dirac equation, all
fermions should have antiparticles, both the elementary and the
composite ones. From this it could follow that antibosons do not
exist. However, the mesons, although bosons with integer spins,
consist of fermions: quarks and antiquarks, so exchanging
quarks and antiquarks in them, we get an antiparticle of the original
meson. Such a pair, for instance, the positive and negative pion, the
bound states $\mrm{\pi^-=[\ol{u}d]}$ and $\mrm{\pi^+=[u\ol{d}]}$,
where u and d are the two lightest quarks and the antiparticles are
denoted by overline. Sometimes the charged bosons,  W$^+$ and W$^-$
are erroneously considered to be antiparticles of each other. 
The frequently mentioned statement that the photon is its own
antiparticle is a complete nonsense.

\section{Negative energies}

Thus we have indeed real antiparticles, but how to interpret their
negative energy implying a negative mass when at rest? One of the
basic laws of physics is $CPT$ invariance. It states that performing
three discreet transformations simultaneously, charge conjugation $C$
turning particles into their antiparticles, space reflection $P$
changing the directions of all three coordinate axes, and time
reversal $T$ should keep the measurable physical quantities
unchanged. This of course implies that particle and its antiparticle
should have exactly the same properties apart from the signs of their
charges, but also that free antiparticles can be treated as
particles going backward in space-time.

\begin{figure}[t]
  \begin{minipage}[c]{0.25\linewidth}
  \includegraphics[width=\linewidth]{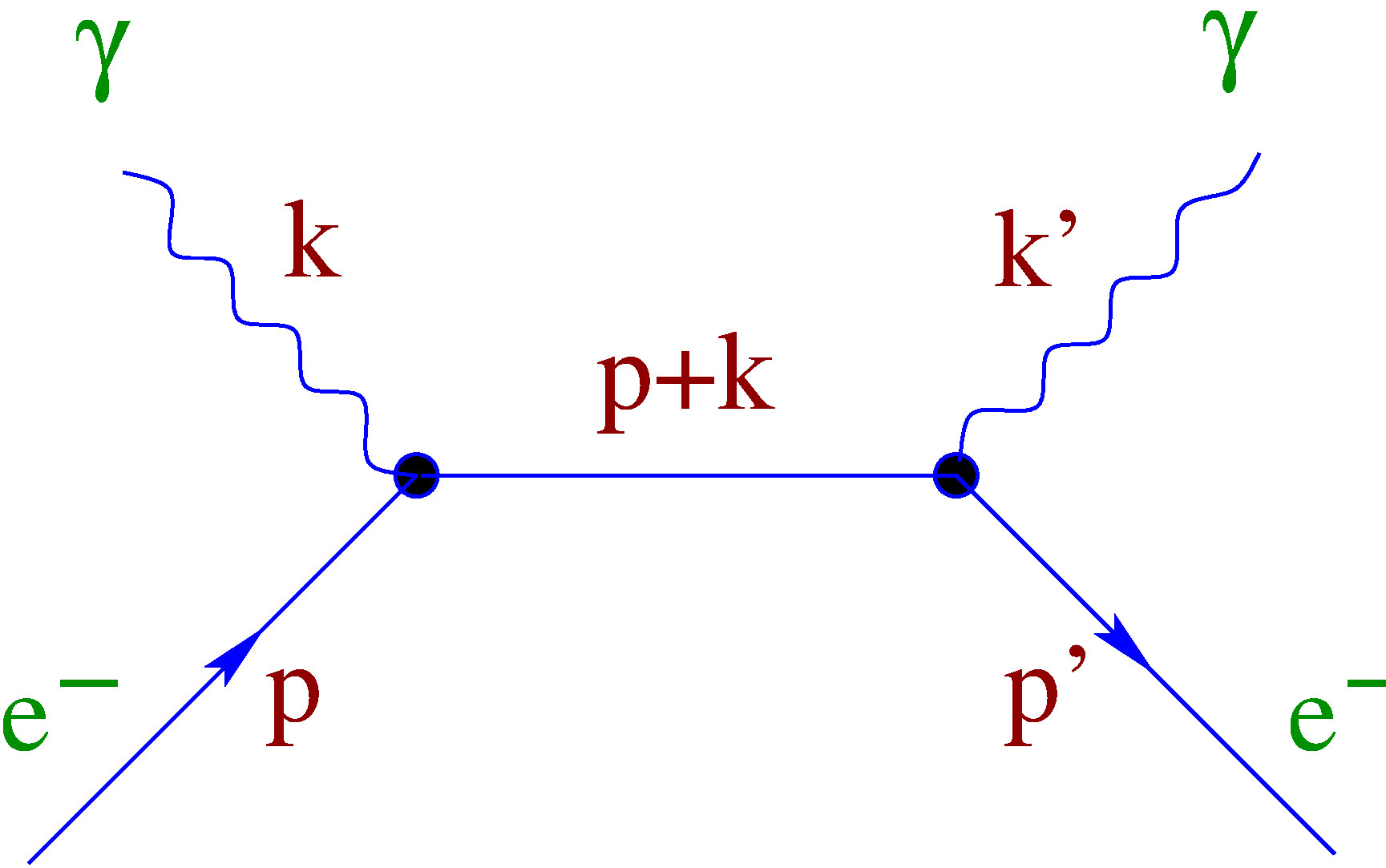}
  \end{minipage}
  \begin{minipage}[c]{0.02\linewidth}
      {\Large +}
   \end{minipage}
  \begin{minipage}[c]{0.25\linewidth}
  \includegraphics[width=\linewidth]{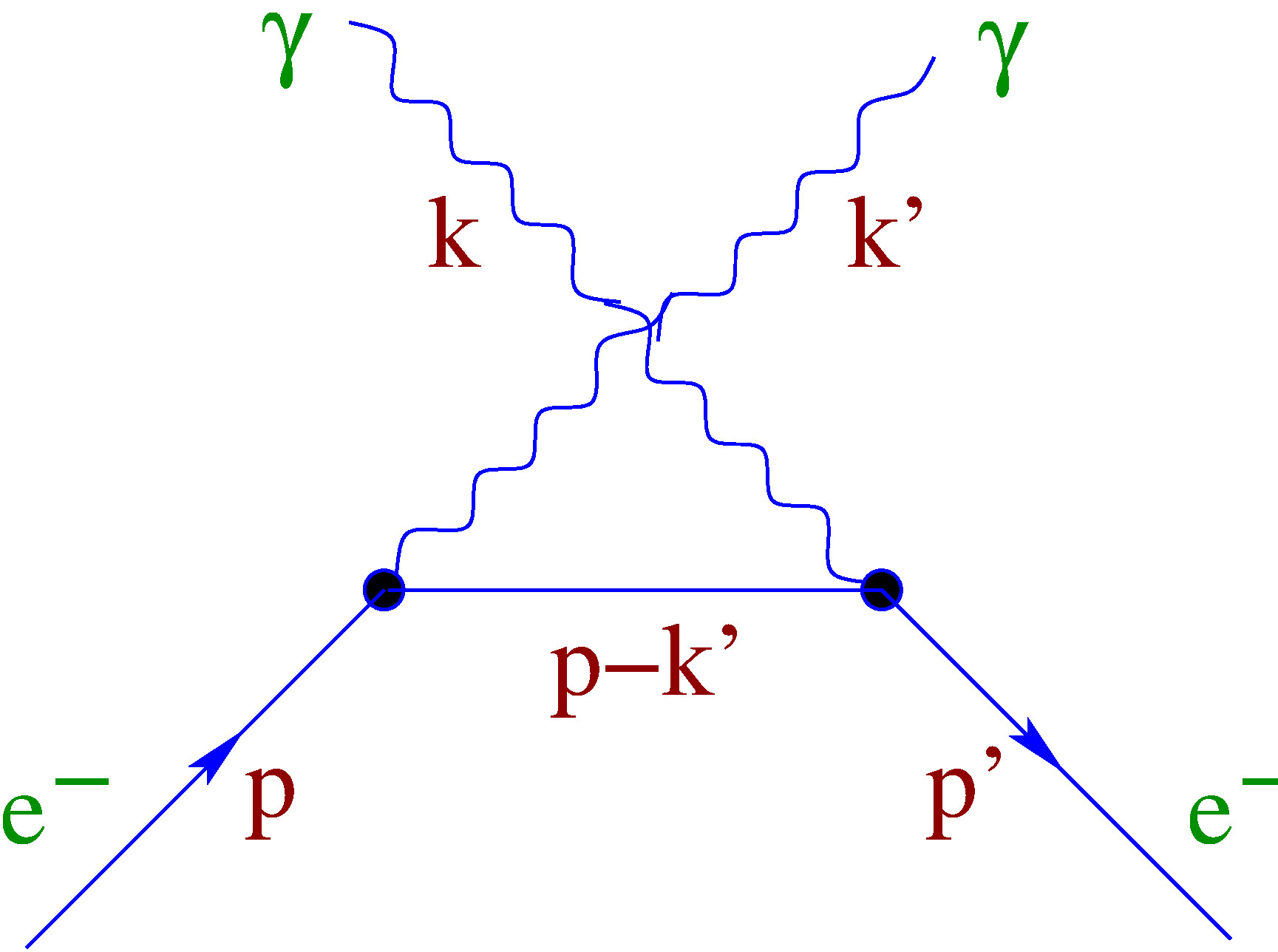}
   \end{minipage}~
  \begin{minipage}[c]{0.1\linewidth}
    {\Large $\Leftrightarrow$}
    \end{minipage}
  ~\begin{minipage}[c]{0.12\linewidth}
  \includegraphics[width=\linewidth]{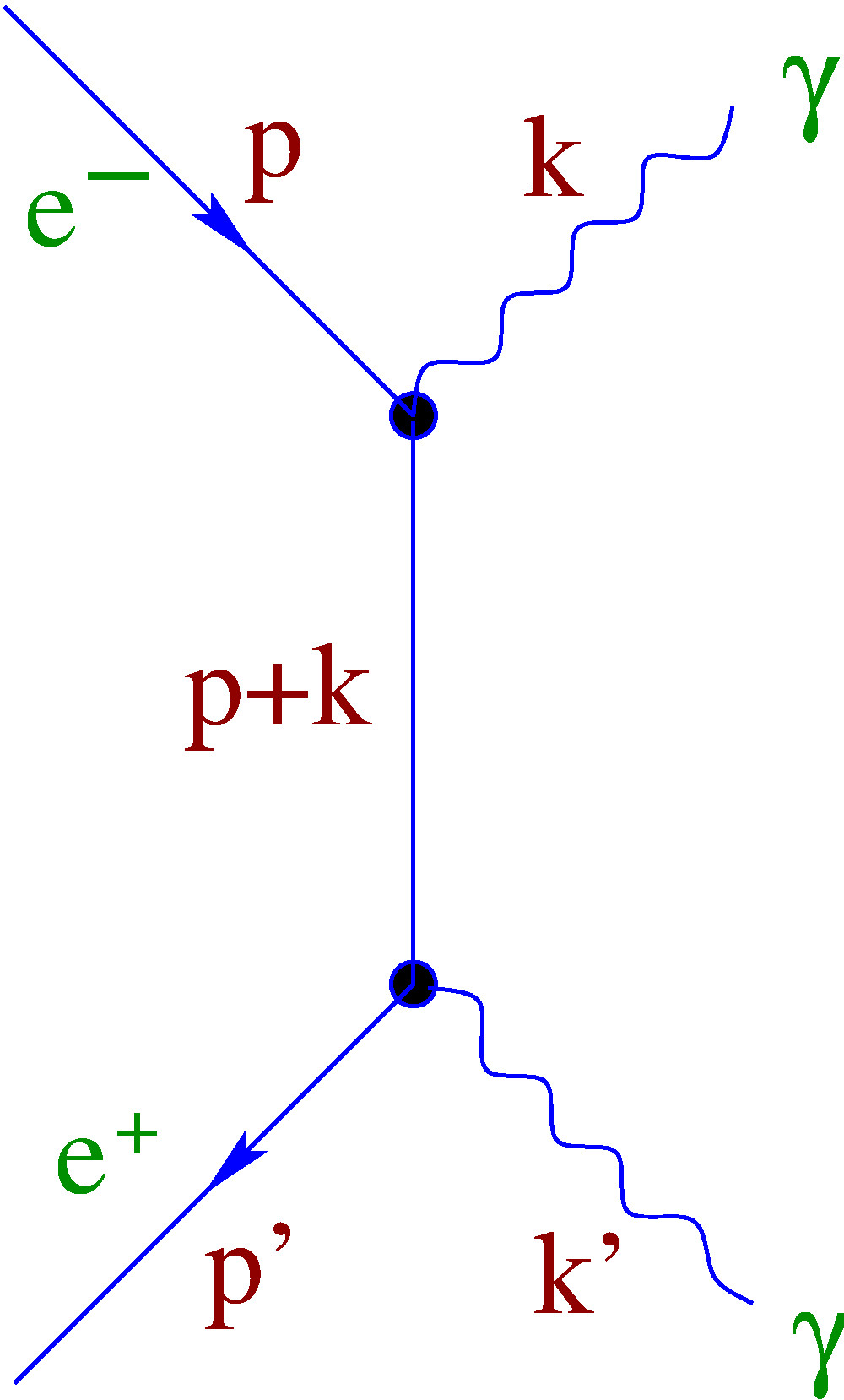}
  \end{minipage}
  \begin{minipage}[c]{0.02\linewidth}
  {\Large +}~
  \end{minipage}
  \begin{minipage}[c]{0.16\linewidth}
  \includegraphics[width=\linewidth]{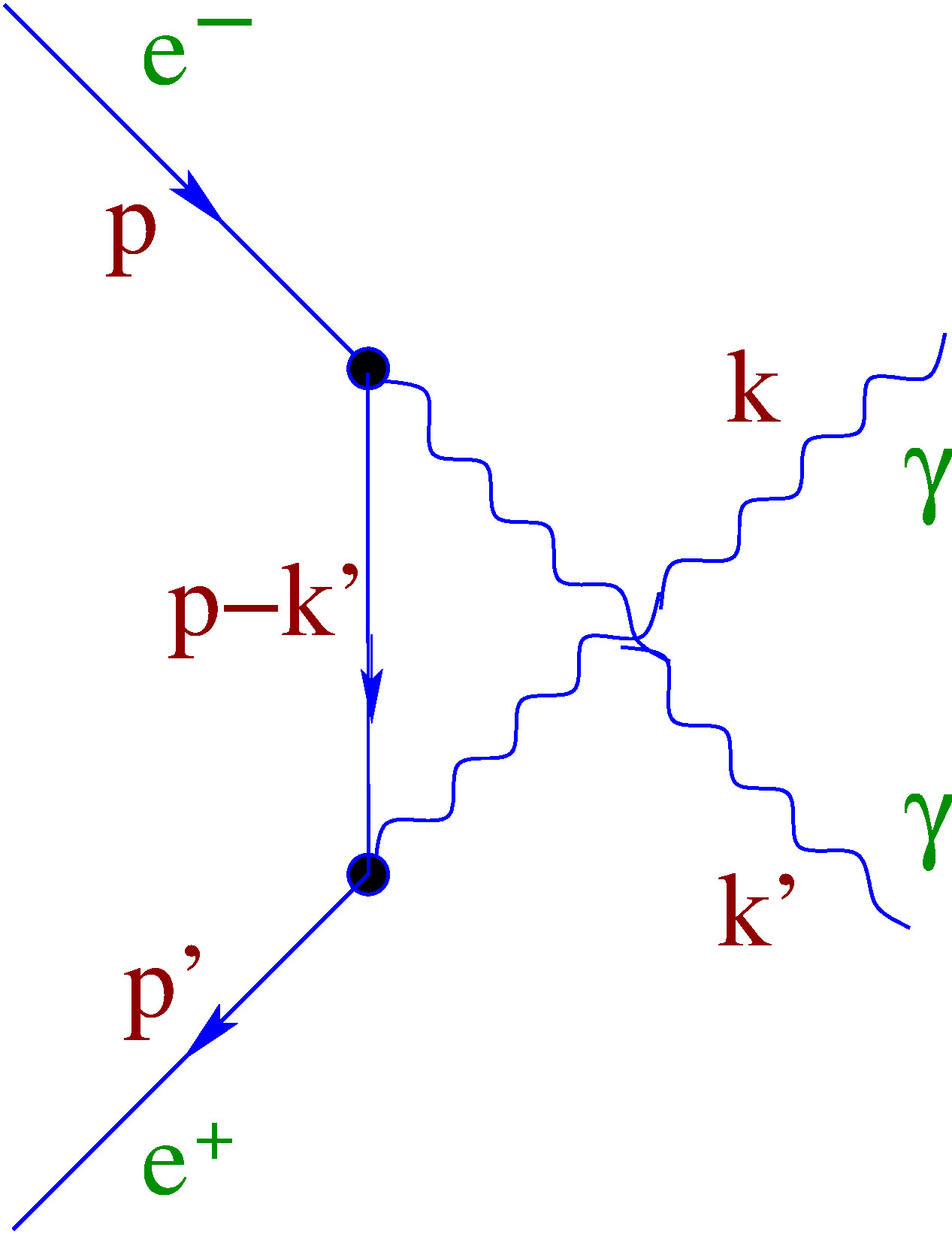} 
  \end{minipage}
  \caption{\footnotesize Photon scattering on electron (left) and
    electron-positron annihilation to a pair of photons (right). The
    photons are interchangeable, that is why we need two amplitude
    terms.
}
\label{fig:Compton-annih}
\end{figure}

The problem with the antiparticle concept was addressed by R.\ P.\
Feynman \cite{Feynman:1949hz} in 1949. Following {\em Ernst Stückelberg}
\cite{Stueckelberg:1941th} he states that Dirac's negative energy
states ``appear in space-time ... as waves traveling away from the
external potential backwards in time''. In fact, this way Dirac's hole
theory and the concept of Dirac sea is replaced by real antiparticles 
propagating with time reversed. This principle is employed by the
Feynman diagram technique, and the practically limitless use of Feynman
diagrams with a complete experimental confirmation of the predictions
is one of the most convincing proofs of $CPT$ invariance. As an
example, let us consider electron-positron annihilation to two photons
(Fig.~\ref{fig:Compton-annih}), which can be described as if an electron
arrived to an external field, emitted two photons and left the
premises backward in space-time. Moreover, using a simple diagram
transformation called crossing, the same technique yields the cross
section of Compton scattering of the electron.

\section{Identical to what extent?}

Does the standard model establish indeed the equivalence of particles
and antiparticles? This seems to be implied by the $CPT$ invariance
principle, but of course for {\em free particles only}. What about
interactions? Particles and antiparticles have the same mass as
demonstrated by the kaon masses: the relative mass difference between
the neutral kaon and anti-kaon as measured using kaon
oscillation~\cite{ParticleDataGroup:2022pth} is less than
$10^{-18}$. Thus the equivalence should be valid for gravity with
energy (or equivalently, mass) as its source. For the strong
interaction the colour--colour, colour--anti-colour, and
anti-colour--anti-colour attractions are the same for all possible
combinations, so it is correct for quantum chromodynamics (QCD) as
well. Electromagnetism (QED) has a twist --- as an only case among
these interactions --- there is repulsion between identical electric
charges, but otherwise different charges behave identically with each
other, so the equivalence stands here as well.

Do particles and antiparticles have the same physical properties with
just opposite charges? As often, the weak interaction is the
exception. As a result of parity violation, the beta decay, like that
of the muons: $\mrm{\mu^+ \ra e^+ \nu_e \ol{\nu}_\mu}$ and $\mrm{\mu^-
  \ra e^- \nu_\mu \ol{\nu}_e}$ produces left-handed (i.e. in the
massless limit left-polarized, with their spin oriented against the
direction of movement) particles and right-handed antiparticles. That
obviously creates a difference between them. One could ask whether
this problem could be resolved by defining the antiparticles with a
$CP$ transformation instead of the simple $C$. However, that does not
help either as the weak interaction violates the $CP$ symmetry as
well. Moreover, because of $CPT$ invariance, $CP$ violation means
breaking the time reversal symmetry as well, as recently shown by
experimental data.

\section{Charges}

Charge is what makes the basic difference between a particle and
its antiparticle, and thus it is important to clarify the concept of
charge when discussing antiparticles. Moreover, the concept of particles
as generally used in atomic and nuclear physics, could lead to a
misconception of charge.

There are many different kinds of charge. We have mentioned already
the colour charge as the source of the strong interaction, which is
carried by the quarks and the gluons, and we define conserved baryon
and lepton charges as well. One of the basic quantities in the
standard model is the weak hypercharge defined as $Y = 2 (Q-T_3)$
where $Q$ is the electric charge and $T_3$ is the third component of
the weak isospin 
(Table~\ref{tab:fermions}).
This is the source of the
local $U(1)$ interaction both leading to electromagnetism and contributing
to the neutral weak currents.

In a quantum theory there are quantum numbers: numbers with physical
meaning, eigenvalues of physics operators. The basic fermion fields
are eigenstates of various charge operators. For instance, the
electron is an eigenstate with non-zero eigenvalue of the operators of
the electric charge (with eigenvalue --1), of the third component $T_3$
of the weak isospin (--\half), of the weak hypercharge $Y$ (--1),
and of the lepton number (+1). Furthermore, it is a trivial eigenstate
with zero eigenvalue of the operators of colour charge $t_i$, of baryon
number $B$, of the third component $I_3$ of the strong isospin and of
all quark flavour operators.

For the interactions their strength is determined by the product of
{\em charge and coupling}, and we usually choose convenient units for the
charges. That is how the electric charge of the electron became
--1. All other charges are determined by experiment, e.g.\ the vector
currents of the quarks couple to the electromagnetic field with
electric charges $+\frac{2}{3}$ and  $-\frac{1}{3}$, those are their
eigenvalues of the electric charge operator acting on them 
(Table~\ref{tab:fermions}).

\begin{figure}[htp]
  \centering
  \includegraphics[width=0.6\linewidth]
      {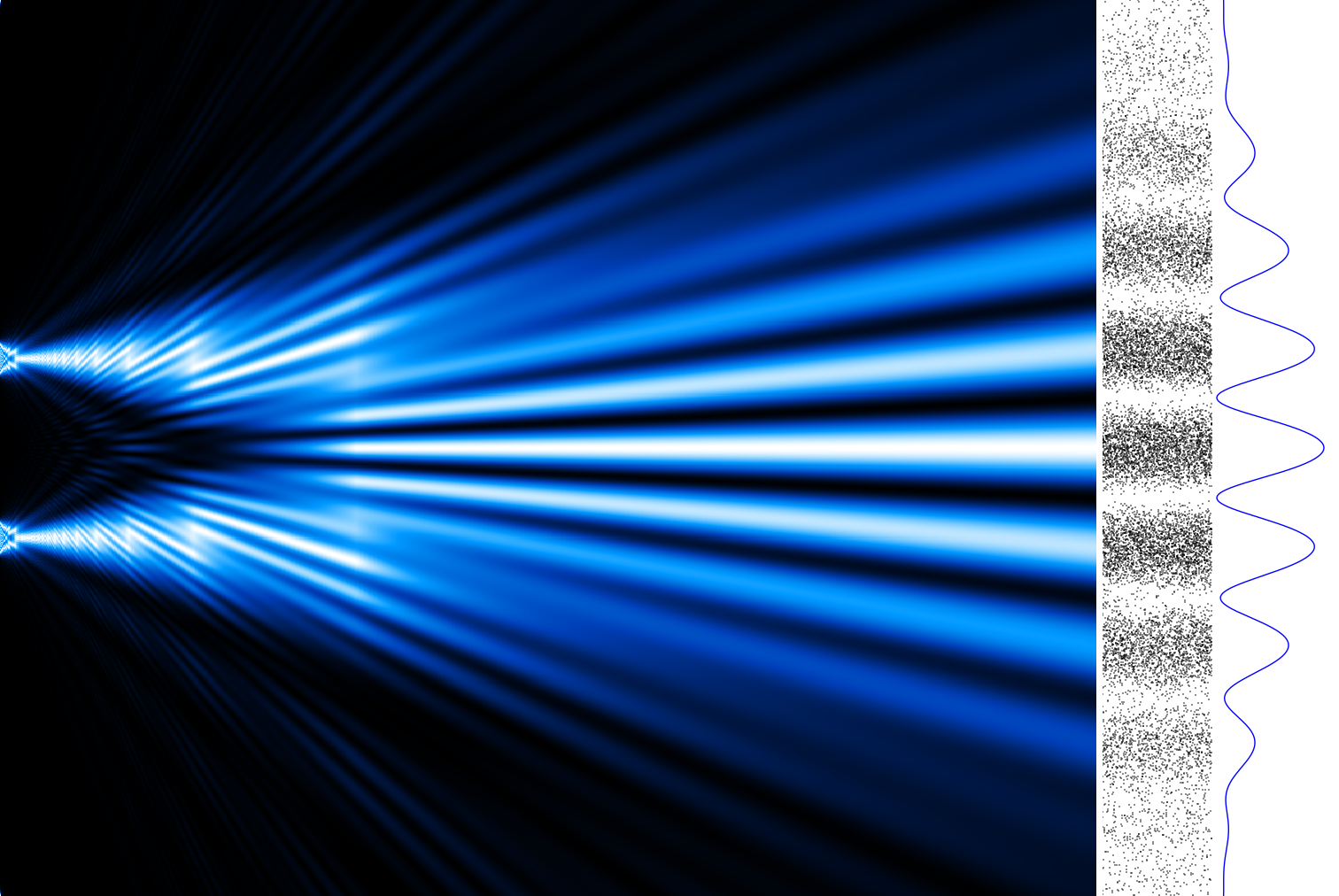}
 \caption{\footnotesize Interference of single electrons traversing two slits \cite{Jonsson:1974}}
\label{fig:double-slit}
\end{figure}

It is well known that one can perform \cite{Jonsson:1974} Young's
double-slit experiment with electrons. On the screen behind the slits
the electrons are detected as point-like particles, so it seems to be
logical to ask, which slit was traversed by the electron. Done with
many consecutive single electrons \cite{Tonomura:1989}, we get a
typical interference picture as with light waves
(Fig.~\ref{fig:double-slit}). According to the generally accepted
interpretation, the electron has both particle and wave forms, and as
a wave passes both slits. We also say that the electron has a charge
$-e$, where $e$ is the elementary charge quantum connected to the
particle-like electron. Returning to the previous question: which slit
was traversed by the charge of the electron? The answer should be
``both'', but that contradicts to the concept of indivisible unit
charge. We have to conclude that the $e$ elementary charge with the
fixed value of $e=1.602076634\cdot 10^{-19}$\,C is just a unit serving
the definition of the electric current measured in Amperes, A = C/s.

This is a paradox resolved by the quantum field theory (QFT) of
particle physics. The electron -- together with all basic fermions,
quarks and leptons -- is a four-component spinor field $\psi(t,x,y,z)$
of the Dirac equation, depending on the coordinates of time $t$ and
space $(x,y,z)$. Of the four components the first two are related to
the two spin states of the particle and the other two to those of its
antiparticle.  In QFT the four-vector $A^\mu = (\phi, \ul{A})$ of the
scalar $\phi$ and vector $\ul{A}$ potential couples with the $j_\mu
=\ol{\psi}\gamma_\mu \psi$ fermion current and thus in the Lagrangian
of quantum electrodynamics this interaction appears as $L_I = e
\sum_{\mu=0}^3 j_\mu A^\mu $. Here $\gamma_\mu$ are the Dirac matrices
and $\ol{\psi}$ is the Dirac-adjoint of the spinor field, while $e$ is
a {\em coupling parameter} characterizing the strength of the
interaction. In field theory such parameters connect (i.e.\ couple)
the particle field current with the external field, generally denoted
by the symbol $g$ (electromagnetism is an exception for historical
reason).

The charge associated with the particle field is a factor
before the $L_I$ interaction term, that is a simple number. Thus it is
meaningless to ask which slit is crossed by the electron charge, just
like to ask where the colour of our eyes traversed the door upon
entering a room. The meaningful question is: where does the electron
go, and the correct answer is through both slits unless we look at
it. The charge is a property of the field, but not a physical
quantity. Nevertheless, the abstraction of conserved charges is a
useful concept in a {\em classical field theory} -- a consequence of
Noether's theorem like the electric charge of electromagnetism --, but
it should not be viewed as physical reality in a {\em quantum theory}.

In this context it is worth to mention that the coupling is not a
constant contrary to the frequent term `coupling constant'. It depends
on the characteristic energy scale of the particle process. The energy
dependence can be predicted precisely in a given quantum field theory.
The meaning of {\em grand unification} is that the couplings of all
fundamental forces become equal at some large energy scale, which
however, does not happen with the couplings of the standard model.

\section{Negative masses?}

The particle masses are also couplings in the Lagrangian, with one
important difference. The $g$ couplings are dimensionless numbers, but
in our particular unit system where $\hbar=1$ and $c=1$ mass has
energy dimension, and so it multiplies directly the bilinear term
$\psi \ol{\psi}$. In the standard model at the fundamental level all
masses are generated by interactions, and as a result the masses also
depend on the energy of interaction between the fields (or particles).

The Dirac equation attributes negative masses to the antiparticles,
which is of course completely virtual, just a mathematical artifact. The
standard model, the generally accepted theory of particle physics, 
is based on quantum field theory which settles 
the problem of negative masses in a simple way: it combines
time reversal with a complex conjugation. As the energy of a particle
at rest carried in the form of a phase factor $e^{\ri Mt}$, time
reversal will keep the mass or energy positive while keeping
everything else unchanged. Thus in the field equations, usually treated
using Feynman diagrams, the antiparticles can be considered as
particles moving in the opposite direction of their momenta. Please
note the use of words: they do not move backwards, just considered to
do that.

In the standard model the masses of the elementary particles are
obtained using the spontaneous symmetry breaking (Brout-Engler-Higgs,
BEH) mechanism. The BEH mechanism creates the masses of the weak gauge
bosons W$^\pm$ and Z$^0$, and allows us to add masses to the fermions
(apart from the neutrinos) using interaction terms of the fermions
with the BEH-field. As these terms have arbitrary couplings, the
obtained masses of the quarks and charged leptons are free parameters
to be determined experimentally. Based on those we select identical
positive values for both particles and their antiparticles. Of course,
all this is related to the basic fermions only, the masses of our
macroscopic world are mostly energy-related, e.g. about 5\,\% of the
nucleon mass is contributed by the three valence quarks, the rest is
kinetic energy contained \cite{Durr:2008zz} in the nucleon. This has
to be emphasized in order to kill the misconception in the general
public raised by newspaper interviews after the 2012 discovery of the
Higgs boson at the LHC.

Nevertheless, the negative masses of the Dirac equation excited the
physicists from the very beginning. Antiprotonic experiments have
shown that the inertial masses of protons and antiprotons are equal
\cite{ParticleDataGroup:2022pth} within the sensitivity of the
measurements, at a relative precision of $10^{-12}$. From the point of
view of gravity, $CPT$ just states that an apple should be attracted
to Earth the same way as an anti-apple to an anti-Earth, the identical
gravitational force of Earth to particles and antiparticles is implied
by the Einstein theory. Nevertheless, that is one of those symmetry
principles which must be tested experimentally. Three experiments are
under development at the Antiproton Decelerator of CERN
\cite{Chardin:2022zrs,Horvath:2019sbh} for such measurements.

\section{Sterile neutrinos?}

Neutrinos are exceptions with many features because of the
peculiarities of the weak force. The standard model ignores this
problem via assuming that neutrinos have no mass. This is one of its
weak points as the discovery of neutrino oscillations implies that
they do have masses.  However, the neutrino masses are so small, at the
limit of observability, that they do not hurt the calculations of the
cross sections of particle reactions.  Moreover, the neutrinos have a
single interaction only, the weak one. Generally, flavour oscillations
are attributed to the co-effect of two interactions with different
eigenstates, so a standard modell neutrino should not oscillate at
all.

Can we accept that the neutrinos have nonzero masses? We must, but
that opens Pandora's box. Massive neutrinos and antineutrinos must
have both left-handed and right-handed states, but the weak
interaction acts on one of them only. The weak isospin doublets of the
standard model have left-handed particles and right-handed
antiparticles only, and so we have to add the singlet states of
right-handed particles and left-handed antiparticles to the Lagrangian. Those among the
quarks will have electromagnetic and strong interactions, but for the
charged leptons only electromagnetism. This means that the
right-handed neutrinos and left-handed antineutrinos will have no
associated charged leptons and cannot participate in any of the three
interactions of the standard model, hence we call them {\em sterile}
neutrinos. This remarkable deficiency of the standard model can be corrected
\cite{Karkkainen:2021tbh} by assuming an additional, very weak interaction.

We see that in spite of the $CPT$ invariance neutrino and antineutrino
could be very different from each other. This problem would be
eliminated if the neutrino were its own antiparticle, i.e.~a
Majorana particle \cite{Elliott:2014iha} named after {\em Ettore
  Majorana}. Nothing prevents that in the standard model, and it
should lead to a peculiar phenomenon: neutrinoless double
$\beta$-decay, when one decay should produce a neutrino and a
consequent one an antineutrino, with the two emissions compensating
each other. Several such reactions are possible \cite{Dolinski:2019nrj} and many
experiments search for them, so far without success.

\section{Dark matter?}

It is well known that about a quarter of all gravitational energy of
our Universe is carried by {\em dark matter} which fills up the
neighbourhood of the galaxies and is indifferent to the
electromagnetic and strong interactions. It amounts to much more than
all other matter contained by the stars, cosmic dust or gas. Such a
particle does not exist in the standard model, and seems to be the
most serious deficiency of the model. It cannot be an ordinary light
neutrino: although there are trillions of neutrinos and their number
is steadily growing due to the activity of the stars, their masses are
very small, and their gravitational contribution can be
neglected. Moreover, light neutrinos fly almost at the speed of light
and cannot make the slowly moving dark matter cloud observed around
the galaxies.

Supersymmetry, one of the most popular extensions of the standard
model, assumes that fermions and bosons exist in pairs of identical
properties, just with different spins. The lightest such neutral
fermion, maybe the supersymmetric partner of the photon or Z boson,
cannot decay further, but could have enough mass to serve as the
particle of the dark matter. Of course, antiparticles create a problem
here, as elementary bosons do not have them. In order to conserve the
necessary degrees of freedom, we should assume that the dark fermion
is a Majorana particle, which also means that they should annihilate
when getting close enough to each other. That also helps to explain
why they do not form stars or black holes.

Another good candidate for the dark particle is the sterile neutrino, if
it exists with a sufficiently large mass. This is another extension of the
standard model, assuming heavy sterile neutrinos with large masses.

\section{Lack of antimatter in the universe}

Another possible deficiency of the standard model could be connected
to cosmology. According to the Big Bang theory, the development of the
Universe had to proceed through a radiation stage, after which
particle-antiparticle pairs were created. However, we do not see
antimatter galaxies. Calculations show \cite{Elliott:2014iha} that
about $6\times10^{-10}$ parts more particles than antiparticles had to
be created in the universe of our horizon to explain this prevalence of
matter.  The standard model is unable to explain the baryon asymmetry
in the Universe.
There are various suggestions for $CPT$ conserving, and more radically,
$CPT$ violating \cite{Kostelecky:2022idz,Kostelecky:2008ts} extensions
of the standard model. There are many experiments studying antimatter
phenomena as well, CERN's Antimatter Factory, the antiproton
decelerator is fully devoted to that, but so far no sign is observed
for any difference between the masses, charges, magnetic moments, or
weights of protons and antiprotons (see, e.g., Refs.\cite{ALPHA:2020rbx,
  BASE:2022yvh}). The KLOE-2 Experiment \cite{KLOE-2:2021ila} is
recently looking for $CPT$ violation in the kaon sector as well.

\section{Summary}

The standard model, the exceptionally successful theory of particle
physics hiccups a bit when treating antiparticles. Moreover, it cannot
account for the masses of the neutrinos, the lack of antimatter, and
the presence of dark matter in the Universe. All those problems will
be solved somehow, by introducing new symmetries or interactions.

The authors are grateful to the referees for their useful
comments on the manuscript, which helped to improve the paper.  DH
acknowledges support by Hungarian research grants OTKA 128786 and
143477. ZT is indebted to Mr Károly Seller for useful discussions.

\flushleft

\end{document}